# Rapid-scan nonlinear time-resolved spectroscopy over arbitrary delay intervals

T. Flöry[1, *], V. Stummer[1], J. Pupeikis[2], B. Willenberg[2], A. Nussbaum-Lapping[2], F. V. A. Camargo[3], M. Barkauskas[4], C. R. Phillips[2], U. Keller[2], G. Cerullo[3,5], A. Pugžlys[1,6], A. Baltuška[1,6]

[1]*Photonics Institute, TU Wien, Vienna, Austria*
[2]*Institute for Quantum Electronics, ETH Zurich, Zurich, Switzerland*
[3]*Istituto di Fotonica e Nanotecnologie-CNR, Piazza Leonardo da Vinci 32, 20133 Milano, Italy*
[4]*Light Conversion Ltd., Vilnius, Lithuania*
[5]*Dipartimento di Fisica, Politecnico di Milano, Piazza Leonardo da Vinci 32, 20133 Milano, Italy*
[6]*Center for Physical Sciences & Technology, Savanoriu Ave. 231 LT-02300 Vilnius, Lithuania.*
*\*Corresponding author: tobias.floery@tuwien.ac.at*

**Femtosecond dual-comb lasers have revolutionized linear Fourier-domain spectroscopy by offering a rapid motion-free, precise and accurate measurement mode with easy registration of the combs beat note in the RF domain. Extensions of this technique found already application for nonlinear time-resolved spectroscopy within the energy limit available from sources operating at the full oscillator repetition rate. Here, we present a technique based on time filtering of femtosecond frequency combs by pulse gating in a laser amplifier. This gives the required boost to the pulse energy and provides the flexibility to engineer pairs of arbitrarily delayed wavelength-tunable pulses for pump-probe techniques. Using a dual-channel millijoule amplifier, we demonstrate programmable generation of both extremely short, fs, and extremely long (>ns) interpulse delays. A predetermined arbitrarily chosen interpulse delay can be directly realized in each successive amplifier shot, eliminating the massive waiting time required to alter the delay setting by means of an optomechanical line or an asynchronous scan of two free-running oscillators. We confirm the versatility of this delay generation method by measuring $\chi^{(2)}$ cross-correlation and $\chi^{(3)}$ multicomponent population recovery kinetics.** © 2022 Optica Publishing Group under the terms of the Optica Open Access Publishing Agreement

https://doi.org/10.1364/OPTICA.457787

Ultrafast pump-probe spectroscopy is a very powerful technique to investigate the dynamics of photoinduced processes in a variety of systems, from (bio)-molecules to solids. It is typically performed in a stroboscopic fashion, whereby at first a pump pulse excites the system and its photoinduced absorption/reflection change is monitored by a time-delayed probe pulse. Typically, the pump and probe pulses are derived from the same laser source, often using nonlinear optical frequency conversion to obtain photon energies matching the energy levels of the samples and a mechanical delay line to control their timing up to a delay of a few nanoseconds [1]. However, many processes, such as ligand rebinding in proteins [2] and charge recombination in photovoltaic devices [3], occur over a multitude of timescales, from fs to μs, which cannot be accessed by mechanical delays. In addition, delay lines require careful alignment to avoid lateral beam drifts and may suffer from spurious signal variations due to a change of the focused beam diameter induced by divergence. For these reasons, there is a high demand for sources of synchronized, high-energy pulses whose delay can be electronically controlled over a wide range without moving optical arms.

Femtosecond dual-comb sources operating at MHz to GHz repetition rate have revolutionized frequency metrology and linear Fourier-domain spectroscopy by offering an interpulse delay scanning mode, known as equivalent time sampling and in specific configurations ASynchronous Optical Sampling (ASOPS) [4,5], both without the need for a mechanically steered optical delay line and its associated shortcomings for large (>ns) interpulse delays [6,7].

While a powerful technique, ASOPS is not suitable for generating a sufficient dynamic range for time-resolved spectroscopy on a broad class of higher order nonlinear susceptibilities. Firstly, the interpulse delay increment is fixed, which does not allow for efficient sampling of complex dynamics extending over multiple timescales. Furthermore, the pulse energy is limited since amplifying a laser oscillator at its full repetition rate implies a high average power (with corresponding technical challenges on the laser source and thermal damage on the sample). The short inter-pulse spacing also limits the scan range, and the measurements are susceptible to the accumulation of long-lived components, e.g. population of triplet states [9]; such artifacts are a typical challenge of 3rd and higher-order nonlinear spectroscopies. Importantly for our further discussion, even if the single-pulse energy at a full oscillator repetition rate is boosted to a level sufficient for parametric frequency conversion and/or for pump-probe signal generation, there will remain numerous problems with the acquisition of kinetics. One prominent problem being the inability to average the nonlinear signal at a fixed delay point and a substantial waiting time before the same delay point can be revisited again. The waiting time grows inversely proportion to the time-resolution step determined by the detuning of the two cavities.

Dividing up the MHz repetition rate of a master oscillator by capturing a selected pulse in a kHz multipass or regenerative amplifier (RA) allows for the generation of fully electronically tunable interpulse delays, as shown in several published schemes [10–12]. The most intuitive approach is to seed two RA cavities from a common oscillator; by either selecting different oscillator pulses or by using different lengths of the amplification window, pulse spacings in the order of the oscillator repetition period can be generated [13]. This method allows the generation of arbitrarily long interpulse delays at the expense of nanosecond resolution. Fully electronically tunable delays over a broad range were demonstrated by Bredenbeck et al. [10], using two electronically synchronized mode-locked oscillators seeding two individually controlled RAs. However, the time resolution was limited to 1.8 ps due to synchronization jitter between the oscillators.

Another interesting development is the (kHz-) Arbitrary Detuning – ASOPS (AD-ASOPS) technique [14–16]. Instead of relying on two synchronized oscillators, two deliberately strongly detuned oscillators with a correspondingly short beat period are used to seed two RAs. The instantaneous frequency of the two oscillators together with a coincidence event are continuously monitored and evaluated. Notably, it is also possible to recover shorter time-delays on a sub-ps scale by post-calibrating the measured results, for example by evaluating the spectral interference fringes of the amplified pulses. The inconvenience, ultimately translated into significantly longer acquisition times, is that the appearance of a particular delay setting within an expected delay range becomes statistical. [10–12]

Here we report a method capable of a deterministic control of time delay scans from fs to ms with femtosecond accuracy, which overcomes most problems arising in both opto-mechanical delay generation and in earlier attempts at electronic delay control and recovery. Our method is enabled by the use of a low-noise, spatially multiplexed, single-cavity, femtosecond, dual-comb oscillator similar to the one presented in [17]. While this can also be accomplished by two synchronized oscillators [18,19], the intrinsic lower noise and experimental setup simplicity of the single-cavity dual-comb enabled us to achieve sub-100-fs jitter at arbitrary delays between the amplified pulses.

Figure 1 shows the conceptual scheme of the system. A dual-comb oscillator (Yb:CaF$_2$, $\lambda$ = 1050 nm) delivers two femtosecond pulse trains with 2.0 W average power each at a repetition rate $f_{\text{rep1,2}} \cong$ 80 MHz and a repetition-rate detuning of $\Delta f_{\text{rep}} \cong$ 500 Hz. This results in a time delay increment between consecutive pulses $\Delta t = 1/f_{\text{rep1}} - 1/f_{\text{rep2}} \approx \Delta f_{\text{rep}}/(f_{\text{rep1,2}})^2 \approx$ 80 fs. A fraction of the oscillator output is directed towards a cross-correlation setup used to detect the temporal overlap of the pulse trains by generating the sum-frequency signal in a β-barium borate (BBO) nonlinear crystal in type I phase matching configuration.

This signal is detected (PDA10A2, 150 MHz bandwidth) to obtain a trigger "start" signal for the timing electronics. The pulses are stretched in a common grating-based stretcher and amplified in two RA cavities (Yb:CaF$_2$, ~600 fs pulse duration, 1mJ pulse energies; further details in supplementary) built as a monolithic block pumped by a single laser diode. The amplified pulses are then recompressed in a compressor employing a single transmission diffraction grating but two separated beam paths. The RAs employ Pockels cells in order to select single oscillator pulses and control the amplification window. A fresh amplification cycle is started with a certain waiting period to select a pulse pair with the desired delay, every time a pulse overlap is detected by the cross-correlation setup. Therefore, the RAs operate at a repetition rate corresponding to the oscillator detuning frequency $f_{RA} = \Delta f_{\text{rep}}$ = 500 Hz.

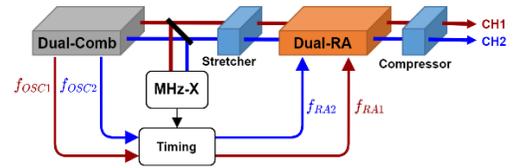

Fig. 1. System setup. MHz-X: cross correlation of the two pulse trains at the full repetition rate for zero overlap detection, Timing: Timing electronics for the RAs.

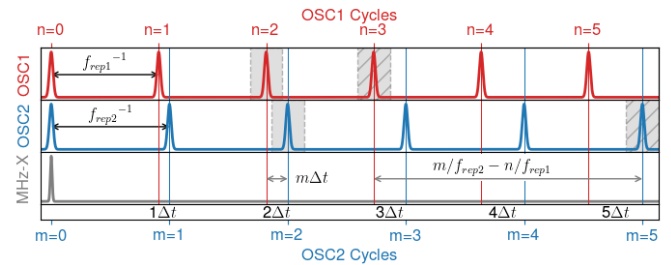

Fig. 2. Operating principle: top trace shows pulses from OSC1, middle trace are the pulses from OSC2, bottom: cross correlation between OSC1 and OSC2. Shaded areas indicate pulses selected for amplification for two cases: gray: $\Delta t < 1/f_{\text{rep1,2}}$, hatched gray: $\Delta t > 1/f_{\text{rep1,2}}$

Figure 2 shows a typical time trace of the pulse train emitted by the dual-comb oscillator. Due to the detuning of the oscillator repetition rates, the pulse delay increases from one pulse pair to the next with a defined increment $\Delta t$. Thereupon, by selecting the pulses of the oscillator pulse trains OSC1 and OSC2, with pulse indices $n$ and $m$, respectively, a pulse pair with a defined time delay $\tau = m/f_{\text{rep2}} - n/f_{\text{rep1}}$ can be selected. For equal pulse indices ($n=m$), this leads to a time delay of $\tau = n(1/f_{\text{rep2}} - 1/f_{\text{rep1}}) \cong n \Delta t$. This allows for the generation of time delays with a step size down to $\Delta t = \Delta f_{\text{rep}}/(f_{\text{rep1,2}})^2 \approx$ 80 fs within a time window from zero optical delay to the inverse oscillator repetition rate $1/f_{\text{rep1,2}} \cong$ 12.5 ns. The latter limitation can be overcome by choosing different waiting times (i.e. pulse indices) for the two channels, e.g. $m = n + k$, so that $\tau = n \Delta t + k/f_{\text{rep2}}$. The pulse indices ($m,n$) and the corresponding interpulse delay are selected by counting the pulses in the respective channel starting from the cross correlation event and then switching the Pockels cells to start the amplification window. Therefore, the interpulse delay can be set on a shot-to-shot basis over the whole range. Furthermore, it is possible to remain at a certain time delay for an arbitrary amount of time by not adjusting the amplification window, allowing to accumulate and average the corresponding transient absorption signal to improve the signal to noise ratio. For interpulse delays in the range of the inverse oscillator roundtrip time the system can also be operated in an ASOPS manner by ignoring the cross-correlation signal and firing the regenerative amplifiers at a fixed repetition rate ($f_{RA}$). Details for this mode can be found in the supplementary.

Single-cavity dual-comb lasers can provide few-femtosecond relative timing jitter [17]. However, at long time scales the relative timing is subject to drifts due to thermal variation in the environment. In our case we have used a single-cavity dual-comb laser for which the repetition rate difference $\Delta f_{rep}$ was stabilized [17]. This stabilization could be avoided by live tracking of the repetition rate difference and recalculating the time delay on a shot-to-shot basis. Using equation 12 from [20] one can estimate the period jitter for the used oscillator to be in the order of 4 fs, which is the determining uncertainty for any pulse pair selected within a 2 ms (=$1/f_{RA} = 1/\Delta f_{rep}$) time window.

As a proof-of-principle demonstration, a cross-correlation measurement of the amplified output pulses was performed. For this measurement the amplified pulses were focused and overlapped, using a single lens, into a BBO crystal and the sum-frequency output signal was recorded using a biased photodiode (Thorlabs DET10A, 350 MHz bandwidth). To suppress scattered light from the fundamental beams, an aperture and optical short-pass filters were installed before the photodiode.

Figure 3 shows the result of the cross-correlation measurements with ~80 fs step size. Panel a) shows the direct result while for panel b) the pulse compressor parameters for one of the channels was changed, by adjusting the optical path length, resulting in a longer output pulse in the corresponding channel and reduced peak intensity.

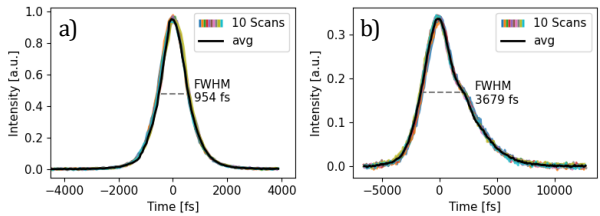

Fig. 3. a) Direct cross-correlation measurements between the two output channels with ~80 fs step size, b) same as a) with one of the compressor channels detuned.

The excellent match between individual scans in Fig. 3 shows the reproducibly of the technique. Furthermore, one can choose for how many cycles the delay value is kept. This allows for example to dwell at a certain delay step for a given number of shots to increase the signal to noise ratio by averaging.

Figure 4a shows the case where for each delay 400 shots were accumulated. In Fig. 4b the intensity distribution for four selected delay values corresponding to the steepest slope of the cross-correlation signal are plotted. Fig 4c shows the corresponding timing jitter as obtained by the slope of the signal from Fig. 4a. Most jitter originates from a uncertainty of the trigger signal which is in the order of one step size ($\Delta t$ = 80 fs). More details on the trigger uncertainty are supplied in the supplementary.

To demonstrate the versatility and flexibility of the presented technique, transient absorption measurements of a perovskite sample (polycrystalline thin film of $CH_3NH_3PbI_3$) were performed using a two color pump-probe setup, which is shown schematically in Fig. 5. Probe pulses with a bandwidth of 15nm were generated in a non-collinear optical parametric amplifier (NOPA), tuned to

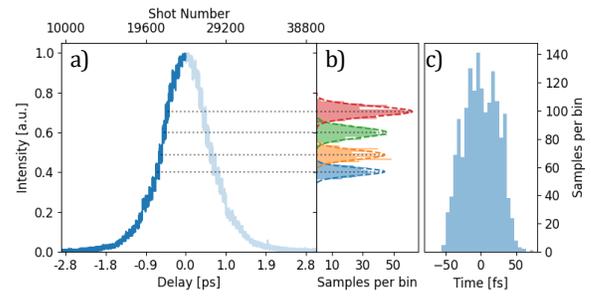

Fig. 4. a) Measured cross-correlation with 400 points per delay. b) Intensity distributions for four delay values c) corresponding timing jitter distribution .

760 nm near the bandgap of the material. The pump pulses at 525 nm were obtained by frequency doubling the fundamental output of the other RA, providing more than 780 meV of excess energy above the bandgap.

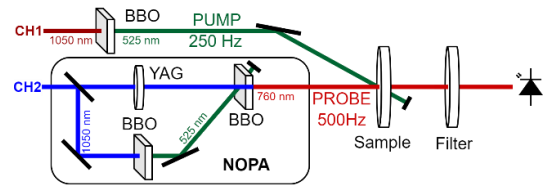

Fig. 5. Two-Color pump-probe setup. One output channel is frequency doubled in a BBO crystal and used as a pump. The second output is frequency converted to 760 nm in a noncollinear optical parametric amplifier. YAG: undoped, for white light seed generation

Figure 6 shows the dynamics of photoinduced transmission changes acquired for delays up to 50 ns with user controllable step sizes varying from 80 fs at early times to 512 ps at longer delays. The transient was recorded using a photodiode and a lock-in amplifier. A total of 463 datapoints were recorded in under 6 minutes. An electronic chopper was implemented by firing the RA for the pump pulses at half the repetition rate (250 Hz). The ultrafast photophysics of $CH_3NH_3PbI_3$ was studied in detail before [21]. The data in Fig. 6 represent the dynamics of the transmission at a probe wavelength of around 760 nm, where the photoinduced absorption due to bandgap renormalization [22], which leads to a red shift of the absorption band, dominates at short delay times. As charge carriers cool down and relay to the band edge, this photoinduced absorption is gradually transformed to the photobleaching characteristic of thermalized excitons and free carriers, causing a change in sign of the transmission changes. Formation of the positive signal takes place within 2 ps, in agreement with previous studies [21,22]. Afterwards, the signal decay on the nanosecond timescale, reflecting a mixture of trap-assisted recombination and Auger recombination the dynamics which are highly dependent on material preparation methods [23,24].

In conclusion, we have demonstrated a versatile scheme for the flexible generation of amplified femtosecond pulse pairs with rapidly variable, electronically tunable delays ranging from femtosecond to milliseconds with femtosecond precision, enabled by a low-noise, spatially multiplexed, single-cavity, femtosecond, dual-comb solid-state oscillator in combination with a twin amplifier. In the demonstrated implementation the timing uncertainty between two arbitrarily delayed amplifier pulses is limited to the delay step $\Delta t = \Delta f_{rep}/(f_{rep1,2})^2$ which in our case was 80 fs at $\Delta f_{rep}$ = 500 Hz rate. However, by pairing such system with a single-cavity dual-comb oscillator at 1 GHz repetition rate [25], 1 fs timing uncertainty at 1 kHz amplifier rate can be achieved.

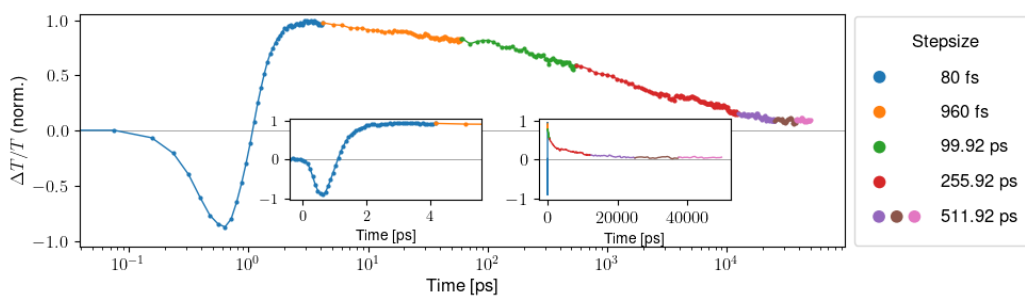

Fig. 6. Pump-probe data of the perovskite sample recorded over 50 ns with varying resolution and logarithmic time axis. The insets show the data with linear time axis in two different time scales. The legend denotes the step sizes used for each timescale.

The presented approach allows the recording of dynamics in nonlinear transient absorption spectroscopies ranging over orders of magnitude, from femtoseconds to milliseconds. We envisage, that, besides the present use in ultrafast spectroscopy, the presented method will also enable several new applications in nonlinear beam steering and time / frequency to space mapping by utilizing presented rapid delay control for time-gating intense chirped pulses.

**Funding.** FWF (4566); Schweizerischer Nationalfonds zur Förderung der Wissenschaftlichen Forschung (40B1-0_203709, 40B2-0_180933).

**Disclosures.** The authors declare no conflicts of interest.

**Acknowledgments.** We acknowledge Patrice Camy (CIMAP) for providing Yb:CaF2 crystal used in the single-cavity dual-comb laser.

**Supplemental document**. See Supplement 1 for supporting content.